%% file: draft_main.tex
\documentclass[12pt]{article}
\pdfoutput=1
\usepackage{hyperref}
\usepackage{amsmath,amssymb,amsthm}
\usepackage{graphicx}
\usepackage[usenames, dvipsnames]{xcolor}
\usepackage{cite}

\newcommand{\be}{\begin{equation}}
\newcommand{\ee}{\end{equation}}
\newcommand{\beq}{\begin{eqnarray}}
\newcommand{\eeq}{\end{eqnarray}}

\newcommand{\cL}{{\cal L}}
\newcommand{\cO}{{\cal O}}

\def\simlt{\stackrel{<}{{}_\sim}}
\def\simgt{\stackrel{>}{{}_\sim}}

 \usepackage{tikz}
\usetikzlibrary{trees}
\usetikzlibrary{decorations.pathmorphing}
\usetikzlibrary{decorations.markings}
\tikzset{
    photon/.style={decorate, decoration={snake}, draw=black},
    wino/.style={draw=redwine},    
    electron/.style={draw=black, postaction={decorate},
        decoration={markings,mark=at position .55 with {\arrow[draw=black]{>}}}},
    scalar/.style={draw=black, dashed,postaction={decorate},
        decoration={markings,mark=at position .55 with {\arrow[draw=black]{>}}}},
    gluon/.style={decorate, draw=black,
        decoration={coil,amplitude=4pt, segment length=5pt}}
}
  
\definecolor{paleblue}{rgb}{0.69, 0.93, 0.93}  
  
\title{\bf Continuum-Mediated Dark Matter--Baryon Scattering}

\author{Andrey Katz$^{a,b}$, Matthew Reece$^c$, and Aqil Sajjad$^c$\\
{\small \texttt{andrey.katz@cern.ch, mreece, sajjad@physics.harvard.edu}}\\
{$^a$ \small \em Theory Division, CERN, CH-1211 Geneva 23, Switzerland}\\
{$^b$ \small \em Universit\'e de Gen\`eve, Department of Theoretical Physics and Center for Astroparticle}\\
{\em \small  Physics (CAP), 24 quai E. Ansermet, CH-1211, Geneva 4, Switzerland}\\
{$^c$ \small \em Department of Physics, Harvard University, Cambridge, MA 02138, USA}}

\begin{document}

\maketitle
\vspace{-11cm}
\begin{flushright}
{\footnotesize CERN-PH-TH-2015-216} 
\end{flushright}
\vspace{8.7cm}

\begin{abstract}
Many models of dark matter scattering with baryons may be treated
either as a simple contact interaction or as the exchange of a light
mediator particle. We study an alternative, in which a continuum of
light mediator states may be exchanged. This could arise, for
instance, from coupling to a sector which is approximately conformal
at the relevant momentum transfer scale. In the non-relativistic
effective theory of dark matter--baryon scattering, which is useful
for parametrizing direct detection signals, the effect of such
continuum mediators is to multiply the amplitude by a function of the
momentum transfer $q$, which in the simplest case is just a power
law. We develop the basic framework and study two examples: the case
where the mediator is a scalar operator coupling to the Higgs portal (which
turns out to be highly constrained) and the case of an antisymmetric tensor operator ${\cal O}_{\mu
  \nu}$ that mixes with the hypercharge field strength and couples to
dark matter tensor currents, which has an interesting viable parameter space. 
We describe the effect of such mediators
on the cross sections and recoil energy spectra that could be observed
in direct detection. 
\end{abstract}

\section{Introduction}
\label{sec:intro}
\input{intro}

\section{The scenario: mass scales and kinematics}
\label{sec:setup}
\input{setup}

\section{Higgs Portal}
\label{sec:higgs}
\input{higgs}

\section{Hypercharge Portals}
\label{sec:hypercharge}
\input{hyper}

\section{Conclusions}
\label{sec:conclusions}
\input{conclusions}

\section*{Acknowledgments}

We thank an anonymous referee for pointing out the importance of supernova constraints on dark photons.
The research of AK  was partially supported by the  
Munich Institute for Astro- and Particle Physics (MIAPP) of the DFG
cluster of excellence ``Origin and Structure of the Universe.''
MR is supported in part by the NSF Grant PHY-1415548.  AK's and MR's
work was 
supported in part by the National Science Foundation under Grant
No.~PHYS-1066293 and the hospitality of the Aspen Center for
Physics. MR would also like to thank the organizers of the 21st
International Summer Institute on Phenomenology of Elementary
Particles and Cosmology (SI2015) near Beijing for providing a
hospitable environment while some of this work was completed. 

\appendix

\section{Factorization example: antisymmetric tensor}
\label{sec:tensor}

Let's consider two different scenarios that involve the dark matter
tensor current operator $\chi^\dagger \sigma^{\mu \nu} \chi$ coupling
to some mediator. The first case is the standard electric or magnetic
dipole moment coupling to the photon: 
\be
{\cal L}_1 = c \chi^c \sigma^{\mu \nu} \chi F_{\mu \nu} + {\rm h.c.}
\ee
In the second case, we couple to some more general operator ${\cal
  O}_{\mu \nu}$ of dimension $d \geq 2$ that mixes with the photon's
field strength: 
\be
{\cal L}_2 = c_1 \chi^c \sigma^{\mu \nu} \chi {\cal O}_{\mu \nu} +
{\rm h.c.} + c_2 {\cal O}_{\mu \nu} F^{\mu \nu}. 
\ee
The question is whether these two Lagrangians can lead to different tensor structures or form factors. We will see that they do not. 

In this computation we will use the following two propagators for the
photon and the tensor field ${\cal O}_{\mu \nu}$: 
\beq
\left<A_\mu(q) A_\nu(-q)\right> \equiv \Pi_{\mu \nu}(q) & = &
\frac{1}{q^2} \left(g_{\mu \nu} - \frac{q_\mu q_\nu}{q^2}\right), \\ 
\left<{\cal O}_{\mu \nu}(q) {\cal O}_{\lambda \sigma}(-q)\right>
\equiv P_{\mu \nu,\lambda \sigma}(q) & \propto & -(-q^2)^{d - 2}
\left[\left(g_{\mu \lambda} g_{\nu \sigma} - 2 g_{\mu \lambda}
    \frac{q_\nu q_\sigma}{q^2} - 2 g_{\nu \sigma} \frac{q_\mu
      q_\lambda}{q^2}\right) - \left( \mu \leftrightarrow \nu \right)
\right]. \nonumber \\ 
\eeq

\subsection{Hypercharge with dipole moments}

Let's first consider the case of ${\cal L}_1$:

\begin{equation}
\begin{tikzpicture}[line width=1.5 pt]
\draw[electron] (-1.0,-1.0)--(-0.5,0.0);
\draw[electron] (-0.5,0.0)--(-1.0,1.0);
\draw[photon] (-0.5,0.0)--(1.5,0.0);
\draw[electron] (2.0,-1.0)--(1.5,0.0);
\draw[electron] (1.5,0.0)--(2.0,1.0);
\node at (-1.25,1.0) {$\chi$};
\node at (2.25,1.0) {$\psi$};
\node at (-0.85,0.0) {$\mu\nu$};
\node at (1.75,0.0) {$\rho$};
\node at (0.5,0.5) {$q \rightarrow$};
\node at (3.0,0.0) {$\propto$};
\node at (6.5,0.0) {$J^{\rm dark}_{\mu \nu} \left(q^\mu \Pi^{\nu
      \rho}(q) - q^\nu \Pi^{\mu \rho}(q)\right) J^{\rm SM}_\rho$}; 
\end{tikzpicture}
\end{equation}

Because the electromagnetic field strength is dotted into the dark matter tensor current, we have one term where the photon propagator is $\Pi_{\nu \rho}$ and one where it is $\Pi_{\mu \rho}$, each multiplying the appropriate momentum from the derivative from $F_{\mu \nu}$. A little simplification reveals that this amplitude dots the dark tensor current and visible electromagnetic current into the object
\be
{\cal I}_{\mu \nu, \rho}(q) \equiv q_\mu \Pi_{\nu \rho}(q) - q_\nu
\Pi_{\mu \rho}(q) = \frac{q_\mu g_{\nu \rho} - q_\nu g_{\mu
    \rho}}{q^2}. 
\ee

\subsection{Antisymmetric tensor mediator}

Now we consider the case of ${\cal L}_2$, where the coupling is to an
antisymmetric tensor field that kinetically mixes with hypercharge: 
\begin{equation}
\begin{tikzpicture}[line width=1.5 pt]
\draw[electron] (-1.0,-1.0)--(-0.5,0.0);
\draw[electron] (-0.5,0.0)--(-1.0,1.0);
\draw[photon] (-0.5,0.0)--(0.98,0.0);
\draw[photon] (1.52,0.0)--(3.0,0.0);
\draw[electron] (3.5,-1.0)--(3.0,0.0);
\draw[electron] (3.0,0.0)--(3.5,1.0);
\node at (1.25,0.0) {\Huge $\otimes$};
\node at (1.25,-0.5) {$\lambda \sigma$};
\node at (-1.25,1.0) {$\chi$};
\node at (3.75,1.0) {$\psi$};
\node at (-0.85,0.0) {$\mu\nu$};
\node at (3.25,0.0) {$\rho$};
\node at (0.5,0.5) {$q \rightarrow$};
\node at (4.25,0.0) {$\propto$};
\node at (8.5,0.0) {$J^{\rm dark}_{\mu \nu}\, P^{\mu \nu, \lambda
    \sigma}(q) \left(q_\lambda \Pi_{\sigma \rho}(q) - q_\sigma
    \Pi_{\lambda \rho}(q)\right) J^{{\rm SM}\, \rho}$}; 
\end{tikzpicture}
\end{equation}
In this case, we first have the propagator of ${\cal O}_{\mu \nu}$ itself, then an insertion which mixes it into electromagnetism, which propagates and couples to the Standard Model current.

In this case, the structure appearing in the middle is
\be
{\cal I}'_{\mu \nu, \rho}(q) \equiv P_{\mu \nu, \lambda \sigma}(q)
\left(q^\lambda \Pi^\sigma_{~\rho}(q) - q^\sigma
  \Pi^{\lambda}_{~\rho}(q)\right). 
\ee
Contracting all of the indices and simplifying, we find that
\be
{\cal I}'_{\mu \nu, \rho}(q) = 2 \left(-q^2\right)^{d - 2} {\cal I}_{\mu \nu, \rho}(q).
\ee
This shows that the general case of antisymmetric tensor mediator
exchange is equivalent to the case of dark matter with dipole moments,
reweighted by constant factors times an appropriate power of $-q^2$
where $q$ is the momentum exchanged between dark matter and the
Standard Model in the scattering process. 

In particular, when the unitarity bound is saturated and $d = 2$,
the operator ${\cal O}_{\mu \nu}$ is the field strength of an abelian
gauge field and this reduces to the usual case of kinetic mixing. In
that case, the $q$ dependence is exactly as for ordinary dipole moment
dark matter.  

\subsection{Comment}

In fact, this result should follow on general grounds. The amplitude
necessarily has the form $J^{\rm dark}_{\mu \nu} I^{\mu\nu\rho}(q)
J^{\rm SM}_\rho$ where $I^{\mu \nu \rho}(q)$ is antisymmetric in $\mu$
and $\nu$. The only possible tensor structure is ${\cal I}_{\mu \nu,
  \rho}(q)$. More generally, exchange of a mediator field will always
produce at most a small finite set of tensor structures coupling a
dark current to a Standard Model current. If an analysis of the full
set of tensor structures has been carried out for the exchange of
ordinary weakly-coupled particles, then the more general exchange of
operators of arbitrary dimension will not lead to new tensor
structures or form factors, but only to new momentum dependence in the
amplitude. 

{\footnotesize 
\bibliography{dmref}
\bibliographystyle{utphys}
}

\end{document}

%% file: intro.tex
Most of the matter in our universe, by mass, is dark matter, but
beyond the fact that it interacts gravitationally, the  nature of dark
matter remains elusive. As experiments dig further into the parameter
space of classic theories of dark matter and continue to find null
results, it is important that we think as broadly as possible about
what dark matter might be and how we might detect it. In this paper,
we will suggest a novel form of interaction between dark matter and
baryons and explore the extent to which it modifies the signals
experiments searching for dark matter might observe.    

In recent years, a much wider variety of possible dark matter models
and phenomenology has begun to be explored. Non-relativistic effective
theories  have systematized the exploration of possible operators
characterizing dark matter--baryon scattering in direct detection
experiments~\cite{Fan:2010gt, Fitzpatrick:2012ix, Fitzpatrick:2012ib,
Anand:2013yka, Peter:2013aha, Gresham:2014vja, Gluscevic:2014vga,
Catena:2014epa, Schneck:2015eqa, Catena:2015uua, Catena:2015vpa,
Kavanagh:2015jma, Gluscevic:2015sqa}, drawing on older work outside
the dark matter context~\cite{Dobrescu:2006au}. The basic operator
approach can be modified in various ways, for instance through
considering dark matter particles inelastically scattering to or from
excited states \cite{TuckerSmith:2001hy, Graham:2010ca,
Barello:2014uda}, dark matter particles with form factors
\cite{Feldstein:2009tr, Chang:2009yt}, dark matter that scatters
through $2 \to 3$ 
processes~\cite{Curtin:2013qsa, Curtin:2014afa}, scattering of dark matter off two nucleons at once \cite{Prezeau:2003sv,Cirigliano:2012pq,Cirigliano:2013zta}, or (a more radical modification) detection not of dark matter itself but of relativistic DM annihilation products \cite{Cherry:2015oca}. Theories containing a large ensemble of
(possibly unstable) dark matter states have been considered
\cite{Kumar:2011iy,Dienes:2011ja, Dienes:2011sa, Dienes:2012cf}, as have theories
in which only a small fraction of dark matter enjoys a richer set of
interactions \cite{Fan:2013yva, Fan:2013tia, McCullough:2013jma,
Fan:2013bea}. All of this theoretical exploration has helped to
broaden our sense of what realistic theories of dark matter can be,
pointing the way to new signatures that can be tested experimentally. 

Our goal in this paper is to explore yet another modification of the
standard picture of how dark matter interacts with other
particles. Specifically, we will study dark matter interactions that
are mediated by generic operators of arbitrary scaling dimension
(consistent with unitarity bounds), which can be thought of as the
exchange of a continuum of light states. Schematically, we would like
to think about Lagrangians of the form 
\beq
{\cal L} = {\bar \chi} \Gamma \chi {\cal O}_{\rm med} + {\bar \psi}
\Gamma \psi {\cal O}_{\rm med}, 
\eeq
where ${\bar \chi} \Gamma \chi$ stands in for any dark matter bilinear
operator and ${\bar \psi} \Gamma \psi$ for a Standard Model bilinear
and ${\cal O}_{\rm med}$ is some ``mediating operator.'' The precise
technical meaning of the statement that a continuum of states is
exchanged is that the momentum-space two-point function $\left<{\cal
O}_{\rm med}(q) {\cal O}_{\rm med}(-q)\right>$ has a branch cut
extending down to momenta well below the threshold momentum exchange
giving rise to a signal in the experiment, $\left|q\right| \ll q_{\rm
exp}$ (and furthermore that the spectral weight is spread out along
the branch cut rather than being concentrated near a single narrow
peak). Cases where ${\cal O}_{\rm med}$ is a single light or heavy
particle are well-studied, but the general case where it represents
the exchange of a continuum of light states has received little
attention. Such couplings to a continuum of states have appeared in
the phenomenological literature in various guises, e.g. in the RS2
setup~\cite{Randall:1999vf} or the literature on
``unparticles''~\cite{Georgi:2007ek, Grinstein:2008qk}, but (perhaps
surprisingly) has 
been mostly absent from explorations of how dark matter can interact
with the Standard Model. 

Let us mention here some related work in the literature. The
possibility that direct detection could proceed through a continuum of
states which would modify the effective nonrelativistic potential
$V(r)$ was briefly discussed in section 2.2 of~\cite{Fan:2010gt}, but
not developed at length. An unusual velocity dependence of dark matter
annihilation from the Sommerfeld effect due to exchange of a continuum
of states was explored in~\cite{Chen:2009ch}; related work neglecting
the Sommerfeld effect appeared in~\cite{Iltan:2012qf}.  Finally, a
related scenario involving a Randall-Sundrum realization of a tower of
light mediators was studied
in~\cite{McDonald:2010iq,McDonald:2010fe,McDonald:2012nc,vonHarling:2012sz,Jaeckel:2014eba}. 

The outline of this paper is as follows. In Sec.~\ref{sec:setup} we
decribe our setup in detail. In particular, we briefly review the CFT 
formalism that we use in our calculations and describe possible
ways to model a mass gap. We show that the direct detection rates
are almost not affected by our assumptions about the modeling of the
mass gap. We also briefly address basic cosmological concerns related
to our scenario. 
In Sec.~\ref{sec:higgs}
we study continuum mediators coupling to the SM Higgs portal
\cite{Silveira:1985rk,Burgess:2000yq,Patt:2006fw,Kim:2006af,Kim:2008pp}. It
is a simple 
illustration of the general idea, but we find that there is not a
viable parameter space for the sort of direct detection signal we are
interested in.  
In Sec.~\ref{sec:hypercharge} we analyze the DM
direct detection and collider constraints in the case
when the DM-nucleus interaction is mediated via an antisymmetric tensor operator which
couples to the SM via the hypercharge
portal~\cite{Holdom:1985ag,Batell:2009vb}. This case realizes
interesting 
continuum-mediated phenomenology in a parameter space compatible with
various constraints. Finally in the 
last section we conclude. Some technical details are relegated to the
appendix.

%% file: setup.tex
\subsection{The basic picture}

The majority of models of dark matter--baryon scattering considered so far take one of two forms: 
\beq
\begin{tikzpicture}[line width=1.5 pt]
\node at (-3.5,-0.3) {Light particle mediator: };
 \draw[electron] (0,0)--(1,0);
 \draw[electron] (1,0)--(1.7,0.5);
 \draw[dashed] (1,0)--(0.9,-0.7);
 \draw[electron] (0.9,-0.7)--(1.9,-0.7);
 \draw[electron] (0.2,-1.2)--(0.9,-0.7);
 \node at (5.0,-0.3) {$ = J_\chi(p,p-q) \frac{1}{q^2} J_{\rm SM}(k,k+q).$};
\end{tikzpicture} \\
\begin{tikzpicture}[line width=1.5 pt]
\node at (-3.4,0) {Pointlike interaction: };
 \draw[electron] (0,0)--(1,0);
 \draw[electron] (1,0)--(1.7,0.5);
 \draw[electron] (0.3,-0.5)--(1,0);
 \draw[electron] (1,0)--(2,0);
 \node at (4.8,0) {$ = J_\chi(p,p-q)J_{\rm SM}(k,k+q).$};
\end{tikzpicture}
\eeq
In position space, these correspond to potentials $V(r) \propto 1/r$
and $V(r) \propto \delta^{(3)}({\vec r})$, respectively. (These are
point-particle idealizations and should be appropriately convolved
with the nuclear form factor and, if it exists, dark matter form
factor.) We could also consider the case of a massive mediator with
mass $m \sim q$, which would correspond to a Yukawa potential
interpolating between these two extremes.  
In this paper, we consider a different scenario, in which we exchange
a continuum of light modes. One way to think of this is as the result
of coupling to states of multiple light particles:  
\beq
\begin{tikzpicture}[line width=1.5 pt]
\node at (-2.6,-0.5) {Continuum mediator: };
 \draw[electron] (0,0)--(1,0);
 \draw[electron] (1,0)--(1.7,0.5);
 \draw [dashed, fill=paleblue] plot [smooth cycle] coordinates {(1,0)
   (0.7,-0.25) (0.6,-0.5) (0.7,-0.75) (1.0,-1.0)  (1.3, -0.75)
   (1.4,-0.5) (1.3,-0.25) }; 
 \draw[electron] (0.3,-1.5)--(1,-1);
 \draw[electron] (1,-1)--(2,-1);
 \node at (5.2,-0.5) {$ = J_\chi(p,p-q) \left(\frac{1}{q^2}\right)^\alpha J_{\rm SM}(k,k+q).$};
\end{tikzpicture}
\eeq
Here the scale-invariant factor of $(1/q^2)^\alpha$ is a stand-in for
more general possible behavior of the intermediate continuum. In the
simplest case, we could consider just a loop of two light,
non-interacting particles. More generally, the continuum could consist
of multiple particles that are themselves interacting, as suggested by
the shaded region in the figure. These interactions could give the
operator ${\cal O}_{\rm med}$ an anomalous dimension, and in the
strong-interaction limit could render any simple particle
interpretation unreliable. 

The examples that we have written above involve a very important {\em factorization} property, which will apply to all of the models that we consider. Namely, the amplitude for dark matter--baryon scattering is a product of three factors:
\begin{itemize}
\item A Standard Model current or ``portal.'' At the microscopic level, this may involve quarks or gluons. In a realistic direct detection calculation, the coupling is to {\em nuclei}, and so this piece of the amplitude in general involves a nuclear form factor describing the way that protons and neutrons are distributed within the nucleus. Such form factors are conveniently calculated with the code of \cite{Anand:2013yka}.
\item A dark matter current. Generally this is taken to be simpler
  than the Standard Model current, since dark matter is treated as an
  elementary particle. However, more generally, dark matter itself
  could have a form factor (see e.g.~\cite{Feldstein:2009tr,Krnjaic:2014xza,Hardy:2014mqa,Hardy:2015boa}). 
\item The propagation of the mediator. In the case of a contact
  interaction, this factor is simply 1. For a massless particle, it is
  $1/q^2$, corresponding to a long-range force. For this paper, we
  will take it to be of the form $1/(q^2)^\alpha$ (motivated by scale
  invariance) or more generally $1/(q^2 - m^2)^\alpha$ (which, as we
  will discuss below, is a crude but useful toy model for a mass
  gap). 
\end{itemize}
Importantly, the propagation of the continuum mediator that we
consider will always simply have the effect of rescaling
well-understood calculations in the literature by simple functions of
$q^2$. It never requires any new nuclear form factors, for
instance. Furthermore, the dark matter and Standard Model currents may
be in nontrivial representations of the Lorentz group, but the effect
of the mediator will always resemble rescaling the result of a contact
interaction by a Lorentz scalar function of $q^2$. In the case where
the mediator has spin and its propagator may involve complicated
tensor structures, this is not entirely obvious. We work out the
details for an antisymmetric tensor mediator that mixes with the
electromagnetic field strength in Appendix \ref{sec:tensor}, showing
that the amplitude is a simple product of known amplitudes for
electric or magnetic dipole moments and a power law in $-q^2$ for a
general scale-invariant mediator. 

\subsection{A cosmological concern}
\label{subsec:BBNconcern}

The simplest realization of continuum exchange is to couple dark
matter and the Standard Model to an operator in an infrared-conformal
sector~\cite{Randall:1999vf, Georgi:2007ek}. However, this is
potentially cosmologically dangerous, because a conformal sector
behaves as dark radiation and is subject to cosmological bounds due to
its effect on the expansion rate of the
universe~\cite{Gubser:1999vj}. These are usually quoted as constraints
on the 
number of ``effective neutrino species,'' $N_{\rm eff}$, from BBN
($\Delta N_{\rm eff} < 1.44$ at 95\% confidence~\cite{Cyburt:2004yc})
and the CMB ($N_{\rm eff} = 3.15 \pm 0.23$ from Planck combined with
other data~\cite{Ade:2015xua}). If dark matter couples to dark
radiation, these constraints can change in various ways depending on
the form of the coupling~\cite{CyrRacine:2012fz, Cyr-Racine:2013fsa,
  Buen-Abad:2015ova, Lesgourgues:2015wza}, but they remain quite
stringent. The safest way to avoid these bounds is if our continuum of
particles does not extend all the way down to zero mass and develops a
mass gap; if the
particles become nonrelativistic before BBN, they are no longer dark
radiation from the viewpoint of cosmological bounds. For instance, our
approximately conformal sector could confine at a scale of at least a
few MeV. This mass means that during BBN, the continuum of modes
behaves as dark matter rather than dark radiation. We can also arrange
that, after acquiring a mass, the continuum states simply decay before
BBN, dumping their energy into the Standard Model plasma. For
instance, if they couple to the Standard Model through dimension-six
operators suppressed by a scale $\Lambda$, they can have a lifetime 
\beq
\tau \sim \frac{\Lambda^4}{m^5} \sim 7 \times 10^{-7}~{\rm
  sec}~\left(\frac{\Lambda}{100~{\rm GeV}}\right)^4
\left(\frac{10~{\rm MeV}}{m}\right)^5, 
\eeq 
easily decaying before BBN (although stable on collider
timescales). More refined estimates can be done for the particular
models we discuss. In the particular case of the hypercharge portal,
we will discuss such estimates further in
Sec.~\ref{subsec:hyperBBN}. It is also possible that deviations from
thermal equilibrium just before BBN could assist in circumventing the
dark radiation bounds even if there is no mass gap (or the mass gap is
much smaller than the BBN scale)~\cite{Reece:2015lch}.   

\subsection{The direct detection scale meets the BBN scale}

Direct detection experiments search for nuclear recoil events. If a
nucleus of mass $m_N$ recoils with kinetic energy $E_R$, the momentum
transfer is $q = \sqrt{2 m_N E_R}$ and the incoming dark matter
velocity must have been at least $v_{\rm min} = \frac{q}{2\mu}$ where
$\mu$ is the dark matter--nucleus reduced mass. A typical dark matter
velocity in the galactic halo is ${\bar v} = 220 {\rm km}/{\rm s}$,
which (taking $\mu = 100~{\rm GeV}$ for reference) can impart at most
a momentum transfer of $q \approx 147~{\rm MeV}$. A low-threshold
nuclear recoil might be taken as, for instance, $E_R = 2~{\rm keV}$
for a silicon nucleus of mass $\approx 26~{\rm
  GeV}$~\cite{Akerib:2010pv}, corresponding to about a 10 MeV momentum 
transfer. Thus, the range of momentum transfers that are relevant for
direct detection might be roughly construed as  
\beq
10~{\rm MeV} \simlt q \simlt 400~{\rm MeV},
\eeq
with the typical momentum transfer of interest in the middle of this
range and the exact details depending on the particular
experiment. LUX, for instance, studies nuclear recoils in xenon
between about 3 keV and 25 keV~\cite{Akerib:2013tjd}, corresponding to
$27~{\rm MeV} \simlt q \simlt 78~{\rm MeV}$. 

From these estimates we see that, if the continuum of modes mediating
scattering acquires a mass gap $m \approx 10$ MeV in order to avoid
dark radiation bounds during BBN, this mass will be a subdominant
correction to the two-point function $\left<O_{\rm med}(q) O_{\rm
    med}(-q)\right>$ at the values of $q$ that are most relevant for
direct detection. The ``mass-gap'' solution to the BBN bound suffers
from a coincidence problem: we have no explanation for why the gap
should fall in the relatively small interval between the temperatures
relevant for BBN and the momenta relevant for direct
detection. (Particular models may offer solutions, but no general
solution is apparent.) But if we {\em assume} that $T_{\rm BBN} \simlt
m \simlt q_{\rm exp}$, we have interesting direct detection
phenomenology while dodging the cosmological bound. If we prefer to
avoid accidental coincidences of mass scales, we can always pursue the
more elaborate nonthermal cosmologies alluded to above to allow for
much smaller $m$. 

\subsection{The conformal limit}

Now that we've argued that we can at least approximately neglect
masses over the range of momenta relevant for direct detection
experiments, let us introduce the formalism we will use throughout
most of the paper, which assumes the conformal field theory limit. To
outline the basic formalism and the assumptions we rely on, we'll
discuss one of the simplest cases we can consider. Namely, ${\cal
  O}_{\rm med}$ is a scalar operator of dimension $d$, coupling to the
Standard Model through the Higgs portal
\cite{Silveira:1985rk,Burgess:2000yq} and to a Dirac fermion dark
matter particle $\chi$ through a scalar bilinear: 
\beq
{\cal L} = \frac{c_h}{\Lambda^{d-2}} H^\dagger H {\cal O} +
\frac{c_0}{\Lambda^{d-1}} {\cal O} {\bar \chi} {\chi}.  
\eeq
(We now drop the subscript ``med'' for convenience.) The first term is
marginal if $d = 2$; we could imagine generating a 
scalar operator with $d \approx 2$ in various ways, including as a
fermion bilinear ${\bar \psi} \psi$ near the edge of the conformal
window in a QCD-like theory \cite{Cohen:1988sq}. For most of the
discussion in this paper, the only information that we need about the
operator ${\cal O}$ is its two-point function, to which we
assign a simple expression in position space. For future use, we also quote the result for an antisymmetric tensor operator:
\beq
\left<{\cal O}(x) {\cal O}(0)\right> & = & \frac{c_{\cal O}}{4\pi^2 \left|x\right|^{2d}}, \\
\left<{\cal O}_{\mu \nu}(x) {\cal O}^{\lambda \sigma}(0)\right> & = & \frac{c_{\cal O}}{4\pi^2 \left|x\right|^{2d}} I_{[\mu}^{~\lambda}(x) I_{\nu]}^{~\sigma}(x), ~{\rm where}~I_{\mu \nu}(x) = g_{\mu \nu} - 2 \frac{x_\mu x_\nu}{x^2}.
\eeq
Unitarity requires $d \geq 1$ for the scalar case and $d \geq 2$ for the antisymmetric tensor. In momentum space the two-point functions acquire extra prefactors:
\beq
\left<{\cal O}(-q) {\cal O}(q)\right> & \equiv & \int d^4 x\, e^{i q \cdot x} \left<{\cal O}(x) {\cal O}(0)\right> = \frac{\Gamma(2-d)}{4^{d-1} \Gamma(d)} c_{\cal O} (-q^2)^{d-2}, \\
\left<{\cal O}_{\mu \nu}(q) {\cal O}^{\lambda \sigma}(-q)\right> & = & -\frac{\Gamma(3-d)}{4^{d-1}\Gamma(d+1)}  c_{\cal O}\left(-q^2\right)^{d-2} \left(g_\mu^{~[\lambda}g_\nu^{~\sigma]} - \frac{2}{q^2} q_{[\mu}q^{[\lambda}g_{\nu]}^{~\sigma]}\right).
\eeq
In both these expressions $c_\cO$ is a normalization constant and we
will further take it to be~1 to simplify the calculations. 
We work in a mostly-minus metric so that the branch cut arises at
physical timelike momentum $q^2 > 0$. Because we will mostly use the
momentum-space expression, one might wonder why we don't assign it the
simple coefficient $c_{\cal O}$ and shift the unpleasant Gamma
functions into the less-used position space answer. The reason is that
we would like smooth behavior of answers near integer dimension: if $d
\approx n + \epsilon$, then 
\beq
(-q^2)^{n - 2 + \epsilon} \approx (-q^2)^{n-2} \left(1 + \epsilon \log(-q^2) + \cdots\right),
\eeq
in which it turns out that the leading term is removed by contact terms and the $\log(-q^2)$ piece is physical. In fact, the $\log(-q^2)$ factor is, for integer dimension operators, the source of the branch cut corresponding to a continuum of physical states. It appears that it would vanish in the $\epsilon \to 0$ limit, but this is compensated by a $\frac{1}{\epsilon}$ from a pole in the prefactor $\Gamma(2-d)$, which we therefore want to keep. (Further discussion of such subtleties may be found in ref.~\cite{Grinstein:2008qk}.)

The next subtlety to discuss is the meaning of the scale $\Lambda$. We could imagine that this is a true cutoff: local field theory begins at the scale $\Lambda$, at which point operators already exist with unusual scaling dimensions, and we are simply given the Lagrangian. But, especially since the values of $\Lambda$ that we can probe will not be far beyond the TeV scale, it seems more likely that the physics is established at some higher energy scale, possibly in terms of weakly-coupled elementary fields, and RG running toward strong coupling leads to the development of large anomalous dimensions and CFT-like behavior. In this case $\Lambda$ may be a combination of other underlying mass scales.

We will not dwell at length on UV completions in this paper, but let's elaborate on this point. Say that the operator ${\cal O}$ has a nontrivial scaling dimension, perhaps $d = 2+\epsilon$. Then we might parametrize its coupling to the Standard Model as $\frac{1}{\Lambda^\epsilon} h^\dagger h {\cal O}$. This is an unusual expression, and especially if $\Lambda$ is a relatively low scale (say, 10 TeV), we should expect that there is dynamics lurking behind it. Perhaps, for example, at some energy $\Lambda_0$ this originated in a dimension-5 interaction of weakly-coupled fields, $\frac{1}{\Lambda_0} h^\dagger h {\bar \psi} \psi$, and then some other interactions at strong coupling drove the operator ${\bar \psi} \psi$ to have a nontrivial scaling dimension. If these interactions became important at a scale $M \ll \Lambda_0$, then we should match ${\bar \psi} \psi \to M^{1-\epsilon} {\cal O}$, and we infer that the effective scale $\Lambda$ suppressing the operator at low energies is $\Lambda = \Lambda_0 (\Lambda_0/M)^{\frac{1-\epsilon}{\epsilon}} \gg \Lambda_0$. Meanwhile, depending on the UV completion, perhaps there were also contact interactions {\em directly} linking the Higgs to dark matter at the scale $\Lambda_0$. The fact that $\Lambda \gg \Lambda_0$ thus raises two concerns:
\begin{itemize}
\item If $M \ll \Lambda_0 \ll \Lambda$, then we either have to make $\Lambda$ quite large (suppressing our signals) or squeeze a lot of dynamics into energies near the weak scale (possibly giving rise to new constraints).
\item If other baryon--dark matter interactions are suppressed by $\Lambda_0$ rather than $\Lambda$, they may give larger effects than those mediated by ${\cal O}$.
\end{itemize}
These concerns suggest to us that we should focus on cases where ${\cal O}$ is an operator of low dimension. If ${\cal O}$ mediates a long-range force, which is to say, if the effective amplitude for scattering mediated by ${\cal O}$ is proportional to a {\em negative} power of $-q^2$, then the powers of $(\Lambda_0/\Lambda)$ that tend to suppress the effects of ${\cal O}$-mediated exchange may be overcome by powers of the large ratio $\Lambda^2/(-q^2) \gg 1$. Precisely because direct detection operates at low momentum transfer, the effects that we are studying may be observable.

Given any effective continuum-mediated model, it would be interesting to follow this logic through more carefully in concrete UV completions. For now, however, we simply take away the general lesson that continuum-mediated scattering is most likely to be observable when the amplitude involves negative powers of $-q^2$, and is potentially subdominant (though this is UV-dependent) to contact interactions when this is not the case.

\subsection{Implications of the mass gap}
\label{sec:massgap}

For a free field, the change from massless to massive is straightforward: we replace the propagator $i/q^2$ with $i/(q^2 - m^2)$. (For higher spin fields there are also modifications to the tensor structure in the numerator.) But for generic operators, there is no universal prescription for how the mass gap appears in the two-point function. The spectral function will be zero below the threshold mass $m$, and nonzero above it, asymptoting to the gapless answer; but the near-threshold behavior could be quite complicated. For confining gauge theories, for instance, we have no analytic tools to compute the precise spectral function. Nonetheless, it is useful to have some simple examples to make more quantitative statements about to what extent a $\sim 10~{\rm MeV}$ mass gap can alter simple scale-invariant predictions for direct detection rates. For direct detection, we are interested in spacelike values $q^2 < 0$, because $q$ is a momentum transfer; this helps to protect our answers from extreme sensitivity to the near threshold behavior, since the threshold and any associated peaks in the spectral function are at $q^2 > 0$.

Let us explore how sensitive the answer is to different models of the mass gap. The simplest conceivable toy model is to replace $(-q^2)^{\alpha} \to (m^2 - q^2)^{\alpha}$ \cite{Fox:2007sy}. Other toy models can come from loops of massive particles or from confining gauge theories or extra dimensions. For simplicity and concreteness, let us focus on the case of dimension 2 scalar operators, for which the CFT two-point function is (up to normalization) simply $\log(-q^2)$. The simple toy model, then, would replace this with $\log(m^2 - q^2)$. One way to obtain a dimension 2 scalar is as a product of two free scalars; in this case the spectral function arises from a loop integral. If we give the free scalar a mass $m_0$, we introduce a branch cut for the two-particle operator at the threshold $m = 2 m_0$, and find that $\Pi(q^2) \propto \int_0^1 dx\, \log(-x(1-x) q^2 + \frac{1}{4}m^2) + 2$. (We shift the result by two so the asymptotics matches $\log(-q^2)$ at $-q^2 \to \infty$.) A third toy model motivated by the RS2 interpretation of the conformal two-point function is to imagine inserting an infrared brane or ``hard wall'' to produce a mass gap, as in RS1 \cite{Randall:1999ee}, or approximately in confining gauge theory at large 't Hooft coupling (see e.g.~\cite{Strassler:2008bv}). We can then compute
\beq
\Pi(q^2) = \frac{J_0(\sqrt{q^2/q_0^2})\log(q^2/q_0^2) - \pi Y_0(\sqrt{q^2/q_0^2})}{J_0(\sqrt{q^2/q_0^2})}.
\eeq
If $x_1 \approx 2.404$ is the first root of $J_0(x)$, then we choose $q_0^2 = m^2/x_1^2$ to obtain a mass threshold $m$. For this model the spectral function is a sum over poles, as in a large-$N$ gauge theory. In a finite-$N$ theory all of these poles acquire a width; we can approximate this effect on the spectral function by studying the large-$N$ answer slightly away from the real axis, effectively smearing out the narrow states: $\rho_\Delta(s) = \frac{1}{2i} (\Pi(s+i\Delta)-\Pi(s-i\Delta))$ \cite{Poggio:1975af,Csaki:2008dt}. An alternative large-$N$ model, perhaps more representative of real QCD (or other theories with small UV 't Hooft coupling), is the digamma function \cite{Blok:1997hs,Shifman:2000jv,Csaki:2008dt}: $\Pi(q^2) = \psi(-q^2 + m^2)$, which has its first pole at $m^2$ and asymptotes to $\log(-q^2)$ for large negative $q^2$.

\begin{figure}
\centering
\includegraphics[width=0.96 \textwidth]{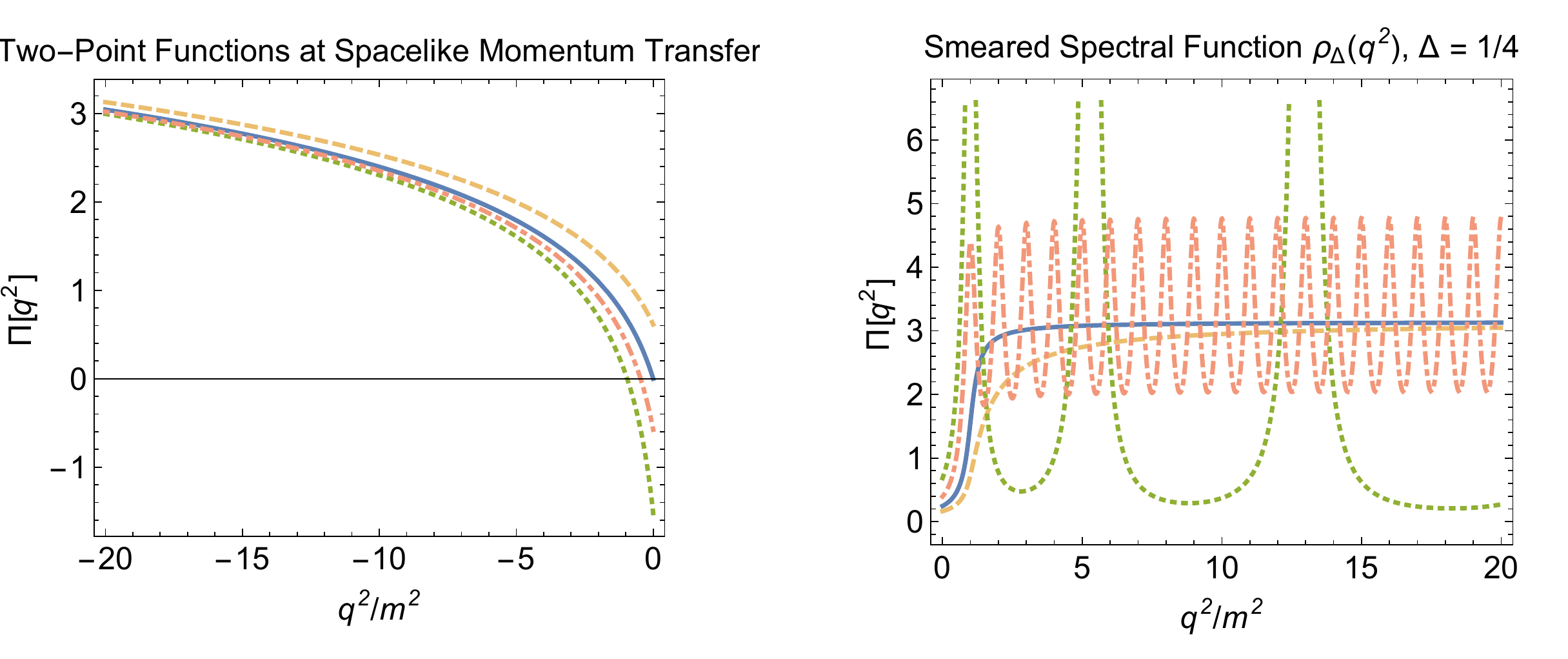}
\caption{Four toy models for a mass gap in the two-point function of a dimension-2 scalar operator. In every case, $\Pi(q^2) \to \log(-q^2)$ in the spacelike region $q^2 \ll 0$. The four models are similar at spacelike momentum (left) but differ greatly for timelike momentum (right). Solid blue: the simple replacement $\log(-q^2 + m^2)$. Dashed orange: the result for a loop of free massive scalars. Dotted green: the Randall-Sundrum or ``hard-wall'' ansatz, characteristic of confinement at large 't Hooft coupling. Dot-dashed red: the digamma function, a toy model of QCD-like confinement.}
\label{fig:massgaps}
\end{figure}

These four toy models are illustrated in Fig.~\ref{fig:massgaps}. The right-hand panel shows the smeared spectral function, which varies from a step function (in the simple $\log(-q^2 + m^2)$ ansatz) to a smooth curve that turns on (the loop of massive scalars) to wildly bumpy curves characteristic of confining large-$N$ theories with many narrow resonances. Despite these dramatically different spectral functions, the behavior for {\em spacelike} momentum, as shown in the left-hand plot, is qualitatively similar in every case. In particular, the corrections to the asymptotic conformal answer when $-q^2 \gg m^2$ are very small. This gives us confidence that the behavior of the direct detection cross section can be approximated with the simplest toy ansatz, $q^2 \to q^2 - m^2$, even without a detailed model for the origin of the mass gap. Such an ansatz will certainly lead to the correct qualitative physics, and will be quantitatively reasonable unless we probe too close to $\left|q^2\right| \approx m^2$. 

Although we have only given examples for the special case $d = 2$, we expect that similar results would be obtained for other dimensions. (Hard and soft wall extra dimensional theories can produce results for arbitrary operator dimension by varying the bulk mass; generalizing the loop ansatz in a well-motivated way appears more difficult.)


%% file: higgs.tex
First, let us briefly analyze the higgs portal for the
continuum-mediated Dirac fermion DM. Because it involves a scalar
operator, this case is formally the simplest. Although we will
immediately see this portal is excluded by the current data, at least
within the 
standard cosmology, some of the results that we get here are relevant
for a much more motivated hypercharge portal.  
The coupling
of the mediator to the SM that we assume in the Higgs portal is 
\beq\label{eq:Hportal}
\cL = c_H\frac{H^\dagger H \cO}{\Lambda^{d-2}}~,
\eeq  
with $d$ being a scaling dimension of the scalar operator $\cO$, whose
critical dimension is~1. The most generic coupling that one could
write down to the fermionic DM is 
\beq\label{eq:ODM}
\cL = c_0 \frac{\cO \bar \chi \chi }{\Lambda^{d-1}} + \tilde c_0
\frac{\cO \bar \chi \gamma^5 \chi }{\Lambda^{d-1}}~.
\eeq  
Note that a priori the scales $\Lambda$ in Eqs.~\eqref{eq:Hportal}
and~\eqref{eq:ODM} can be very different from one another. We absorb
these differences in the couplings $c_0$ and $c_H$ which are not
necessarily $\cO(1)$. 

As we claimed in Sec.~\ref{sec:setup} these interactions can be reduced
to the standard fermionic higgs portal with a $q^2$-dependent
coefficient. This approach is very similar to the approach that we are
taking in the hypercharge portal scenario, where we will define
$q^2$-dependent moments.  The effective higgs portal is 
\beq
\cL = c_{\rm eff}(q^2) \frac{|H|^2 \bar \chi \chi }{\Lambda} + \tilde
c_{\rm eff}(q^2) \frac{|H|^2 \bar \chi \gamma^5 \chi}{\Lambda}
\eeq 
with 
\beq
c_{\rm eff} \equiv \frac{c_H c_0  \Gamma(2-d))}{4^{d-1}\Gamma(d)}
\left( \frac{-q^2}{\Lambda^2} \right)^{d-2}~.
\eeq
The expression for $\tilde c_{\rm eff}$ is identical to that for
$c_{\rm eff}$ up to an obvious change $c_0 \to \tilde c_0$. It is very
easy to qualitatively understand what would be the implications of
this kind of higgs portal on direct detection. The
momentum-dependent coupling $c_{\rm eff}(q^2)$ would translate into an
effective dependence of the recoil spectrum on $q^2$, inducing an
effective form factor, which has nothing to do with either nuclear or dark matter form factors, but rather arises because of the unusual properties of the
mediator. 

Unfortunately, higgs portal continuum-mediated DM is not viable
because of the immediate clash between the induced mass-gap in the
mediator sector and the cosmological constraints. Interestingly, for
any $d <4$, Eq.~\eqref{eq:Hportal} induces a mass gap in the mediator
sector which is 
\beq\label{eq:higgsgap}
m_{\rm gap} = \left( \frac{c_H
    v^2}{\Lambda^{d-2}}\right)^{\frac{1}{(4-d)}}~.
\eeq
Of course if this mass gap is too small we can always assume extra
sources, which might generate an additional mass gap. However, it is
difficult to see how one would significantly reduce this induced mass
gap without very severe fine-tuning. At first glance the automatic presence of a mass gap is {\em good} from the point of view of satisfying BBN constraints; the problem is that the model is too predictive, implying too long a lifetime for a given mass gap.

As we will immediately see the induced mass gap~\eqref{eq:higgsgap} is
too large in those regions of parameter space where we can satisfy
the cosmological constraints. As explained in Sec.~\ref{sec:setup}, the
lightest particle in the mediation sector with a mass of order
$m_{\rm gap}$ should decay faster than one second. Assuming that this
particle is a scalar $s$ with a mass $m_s$ we can write down its
matrix element as  
\beq
\langle 0|\cO | p  \rangle = \xi_s m_s^{d-1} e^{i p \cdot x}~. 
\eeq 
In this expression $\xi_s$ is an unkown $\cO(1)$ constant, that we
will take to one in our further estimates. Therefore the
term~\eqref{eq:Hportal} induces mixing between the scalar $s$ and the physical $h$, which the scalar
$s$ can decay through. Kinematically, only decays $ s \to e^+ e^-$
and $s \to \gamma \gamma $ are allowed and their rates are 
\beq
\Gamma(s \to e^+ e^-) & = & \frac{2 c_H^2 v^2 m_s^{2d-2}}{\Lambda^{2d-4}
  m_h^4} \Gamma_{h\to e^+ e^-} (m_s)\\
\Gamma(s \to \gamma \gamma) & = & \frac{2 c_H^2 v^2
  m_s^{2d-2}}{\Lambda^{2d-4} 
  m_h^4} \Gamma_{h\to \gamma \gamma} (m_s)~, 
\eeq
where $\Gamma_{h \to XX}(m_s)$ stands for the higgs
partial decay widths \emph{if it had a mass $m_s$, rather than
  $m_h$}. In the relevant range of $s$ masses, $1~{\rm MeV} < m_s
< 100$~MeV one can easily see that the $\gamma \gamma $ channel is
completely subdominant and disregard it. The decay to the electron
pairs is the only viable decay channel of $s$, and because the rate is
suppressed by the electron Yukawa squared, the constraints are not
very easy to satisfy. In practice the constraint $\tau_s < 1$~sec
translates to the following bound:
\beq
c_H^2 \left( \frac{\Lambda}{100~{\rm GeV}}\right)^2
\left(\frac{m_s}{10~{\rm MeV}}  \right) \left( \frac{m_s}{\Lambda}
\right)^{2d-2} > 2 \times 10^{-11}~. 
\eeq
It is easy to see that one cannot satisfy this constraint and the
demand $m_{\rm gap} \lesssim 50$~MeV simultaneously. The latter suggests
that $d > 3 $ for any phenomenologically interesting range of the scale
$\Lambda$. The former demands that $d \lesssim 2.34$ for $\Lambda \sim
1$~TeV, and even slightly smaller values of $d$ for higher values of
$\Lambda$. 

Of course, the constraint on the decay time of the lightest narrow
state can be circumvented if we assume non-standard cosmology, but we
still view the constraints on this scenario as not appealing and
further concentrate on the hypercharge portal, which is much more
promising.\footnote{Another potential way to circumvent the
constraints on the higgs portal would be speeding up the decay of the
scalar $s$ by introducing new couplings, which do not necessarily have
anything to do with the DM, {\em e.g.}~$\cO F_{\mu \nu}^2$. However
this is not the most minimal scenario and these types of
couplings come with their own constraints. We relegate the analysis of
these non-minimal scenarios to future studies.} 
Precisely because it does not involve a scalar operator,
the coupling to the Standard Model in the hypercharge portal case does
not automatically deform the CFT and produce a mass gap, so masses and
lifetimes are no longer closely linked and there will be a larger
viable parameter space to explore.

%% file: hyper.tex
\subsection{Portal and recoil spectrum}

We start from studying the hypercharge portal where the hypercharge
field strength couples to an antisymmetric operator,
\beq\label{eq:Ycoupling}
\cL = \frac{c_B}{\Lambda^{d-2}} B_{\mu \nu} \cO^{\mu \nu}~. 
\eeq
The critical dimension of the antisymmetric tensor is 2, so the
dimension of this operator is always bigger than or equal to 4. This is  a
unique portal in the SM, because it allows coupling 
of the BSM bosonic particles to the SM with an operator of
dimension~4.\footnote{Another unique portal which allows coupling 
of the BSM fermions to the SM with an operator of dimension~4 is 
the so-called ``neutrino portal'' $HL$. We will not consider the
fermionic mediators in this paper, mainly because it is not easy to
think
about a scenario, where it would not spoil observed neutrino
properties. } 
Moreover, in the context of the DM, it is the only low-dimension
operator which allows couplings of the DM to the SM which are not
suppressed by the masses of heavy mediators: EW bosons, Higgs or BSM
particles (e.g. $Z'$). 

\begin{figure}[h]
\centering
\includegraphics[width=\textwidth]{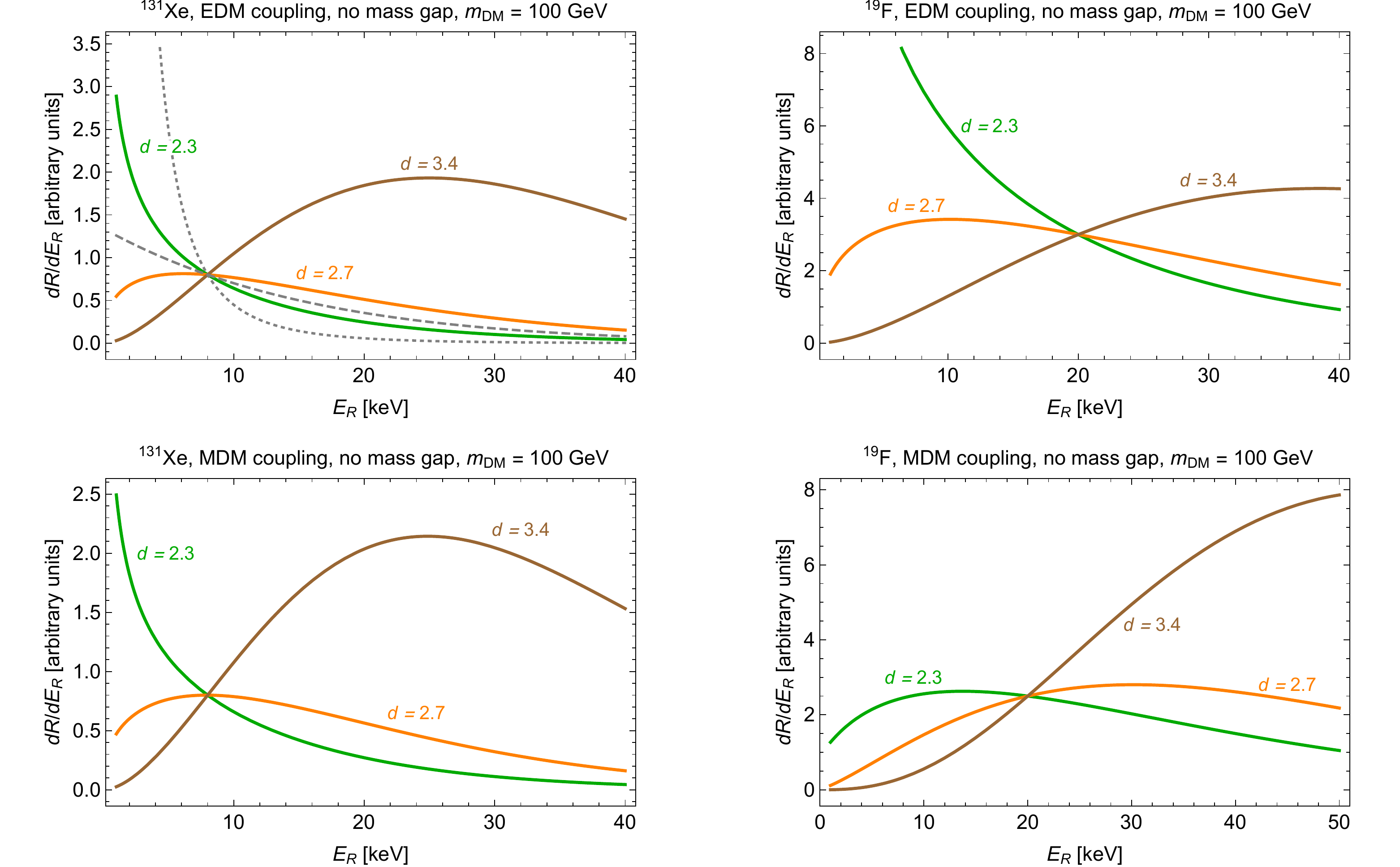}
\caption{Event rate in arbitrary units for electric-hypercharge (top panel)
and magnetic-hypercharge (bottom panel) portals for
different values of $d$. 
Green, orange and brown lines stand for $d
=2.3,\ 2.7,\ 3.4$ respectively.  The curves have been artificially
scaled to intersect at a single point, because the goal is to convey
shape information only. 
The isotope $^{131}$Xe is shown at left and $^{19}$F at right. Notice
that, due to different behavior of nuclear form factors, the shapes
can be quite different for different atoms. 
To guide the reader's intuition for how these compare to more familiar scenarios, in the upper-left figure we also show the recoil spectrum arising from the standard contact operator coupling to mass (${\bar \chi}\chi {\bar N}N$) as a dashed gray line and the same operator with a massless mediator (${\bar \chi}\chi {\bar N}N/q^2$) as a dotted gray line.} 
\label{fig:DMnoGap}
\end{figure} 

The event rate and the total cross sections in the direct detection
experiment strongly depend on the coupling of the antisymmetric tensor
$\cO_{\mu \nu}$ to the DM particles. Hereafter we will only consider
operators with dimensions not higher than~6 in the limit $d [\cO_{\mu
  \nu}] \to 2$.  With this restriction, and assuming fermionic dark
matter one 
can straightforwardly write down three different coupling of the DM to
this operator:
\beq\label{eq:MDMO}
\cL & = & \frac{c_2}{\Lambda^{d-1}} \cO_{\mu \nu} \bar \chi \sigma^{\mu
  \nu} \chi \\
\cL & = & \frac{\tilde c_2}{\Lambda^{d-1}} \cO_{\mu \nu} \chi
\sigma^{\mu \nu } \gamma^5 \chi \\
\cL & = & \frac{\bar c_2}{\Lambda^{d}} \partial^\nu \cO_{\mu \nu}
\bar \chi \gamma^\mu \gamma^5 \chi  
\eeq   
Because of the structure of these operators, which are identical to
the 
magnetic dipole moment (MDM), the  electric dipole moment (EDM) and to
the anapole 
by replacing
$\cO_{\mu \nu} \to F_{\mu \nu}$, we will loosely refer to each of
these options as magnetic, electric and anapole hypercharge portals
respectively. As in the previous sections, we do not necessarily
assume that the couplings $c_B, c_2, \tilde c_2$ and $\bar c_2$ are
order one.  

As we briefly outline in Sec~\ref{sec:setup} and prove in
Appendix~\ref{sec:tensor}, one can 
reduce each of these couplings to the effective dipole/anapole
coupling of the DM to the SM photon, reweighted by an appropriate
power of $-q^2$, where 
the latter is the momentum exchanged between the DM and the nucleus in
the scattering process. Therefore, each of these terms yields an
effective (momentum-dependent) DM dipole moment, which depends on an
appropriate power of 
$-q^2$. The explicit calculation gives the following effective magnetic dipole
moment for the operator~\eqref{eq:MDMO}:
\beq
\mu_{\rm mag} = \frac{c_B c_2  \cos \theta_W }{\Lambda^{2d-3}}
\frac{\Gamma(3-d)}{4^{d-2} \Gamma(d+1)} \left( -q^2 \right)^{d-2} 
\eeq 
For the EDM and the anapole hypercharge portals we get exactly the
same expression for 
the effective moment with an obvious replacement $c_2
\to \tilde c_2$ for the EDM and $c_2 \to \bar c_2, \Lambda^{2d-3} \to
\Lambda^{2d-2}$ for the anapole moment. 

Because all the effective moments depend in a non-trivial way on the
momentum transfer $q^2$, we expect that the differential event rates
$dR/dE_R$ will be modified compared to the ``standard'' DM
dipole/anapole scenario. One can refer to  the factor 
$(q^2)^{d-2}$ as an effective DM form factor. However this form factor
arises because of
the non-trivial dynamics in the mediation sector, rather than
non-trivial structure of the DM. Of course these
effective DM form-factors multiply the nuclear form-factor in the
direct detection picture, yielding distinctive event distributions in
the direct detection experiments. 

It is also worth noticing that the effective DM form-factors will be
further modified compared to the perfect CFT limit by the mass gap
effects. If this mass gap is not too far away from the BBN limit,
namely $m_{\rm gap} \sim \cO(1\ldots 10)$~MeV, this modification is mostly
important in the low-energy recoil
regime. Therefore we expect stronger modifications for the lighter
elements, 
e.g. fluorine and oxygen, while the heavy elements like xenon might
be only very weakly sensitive to the mass gap effects. 

\begin{figure}[t]
\centering
\includegraphics[width=\textwidth]{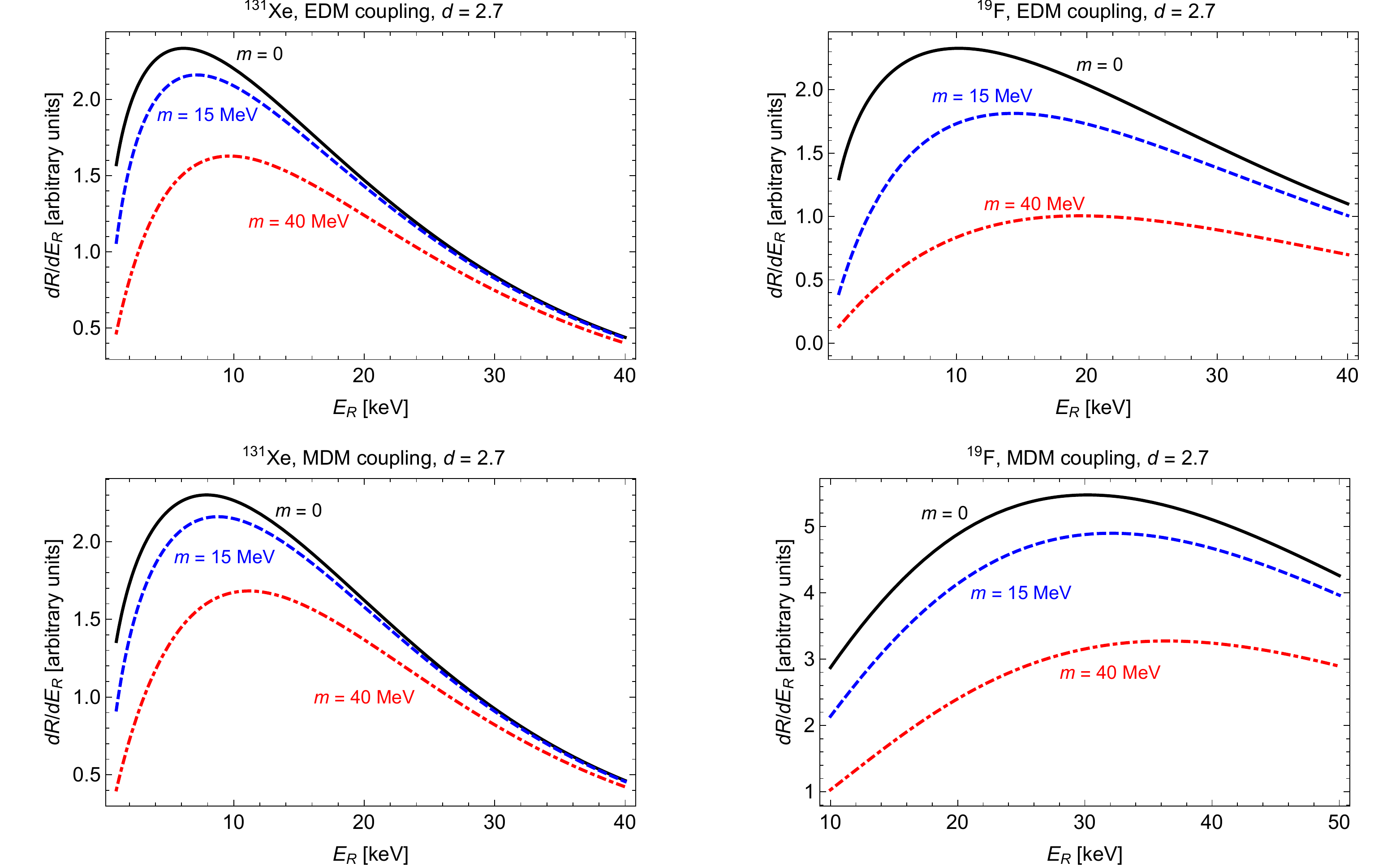}
\caption{Event rate in arbitrary units for electric-hypercharge (top panel)
and magnetic-hypercharge (bottom panel) portals with and without mass gap. Black
(solid), blue(dashed) and red (dot-dashed) lines stand for $m_{\rm gap} =
0,\ 15,\ 40$~MeV respectively. We assume $d = 2.7$ on all these
plots. }
\label{fig:DMd27}
\end{figure}

To illustrate these points we show the expected rate distributions
$dR/dE_R$ assuming scattering on both $^{131}$Xe and $^{19}$F for both electric and
magnetic dipoles on Fig.~\ref{fig:DMnoGap}.\footnote{In xenon
  experiments like LUX and Xenon100 one of course is sensitive to the
  nuclear responses of all
  the seven stable xenon isotopes, and not only to $^{131}$Xe. We checked
  this explicitly, and the distributions that we show do not change
  significantly when all other isotopes are included. Therefore here
  for the illustration purposes we just use one isotope. } 
These figures are meant to
illustrate only the shape of the curve, and so have arbitrary
normalization. The differences between the distributions are 
clearly seen. We further show the rates if the conformal invariance is
broken by mass gaps $m_{\rm gap} = 15$~MeV and $m_{\rm gap} = 40$~MeV in
Fig.~\ref{fig:DMd27}. In this case, the overall normalization remains
arbitrary but the relative normalization of curves on the same plot is
important, showing how the mass gap suppresses the overall rate in a
momentum-dependent manner. To model the mass gap here we use the most
naive model, namely we replace $-q^2 \to (-q^2+m_{\rm
  gap}^2)$. Because the momentum 
transfer is spacelike we assume very little possible difference between
different ways to model this effect (see Sec.~\ref{sec:massgap} for
details). As we   
see the effect is minor or very minor for scattering on $^{131}$Xe,
however it is as expected much more pronounced for scattering on
$^{19}$F.  

From here we can immediately calculate the
total cross sections and the differential event rates in the direct
detection experiments. Of course, there are  differences
between the electric dipole on one side and magnetic
dipole/anapole on another side. While the cross section of the former
is 
enhanced by $q^{-1}$ in the $d[\cO_{\mu \nu}] \to 2$ limit, the cross
sections of the anapole are completely regular and behave as $q^0$ in
this limit. The MDM cross section is technically also enhanced
by $q^{-1}$ and the total cross-section diverges in the deep IR, this
term is suppressed by the transverse velocity, translating to the
non-relativistic operator $\frac{i}{q^2} \vec S_\chi \cdot (\vec q \times \vec
v)$. Therefore, we expect to get much bigger cross sections in
the EDM case than in the other cases for the same values of $\Lambda $
and $d$. The cross sections have been calculated in multiple previous
works  on the dipole/anapole
DM~\cite{Pospelov:2000bq,Sigurdson:2004zp,Barger:2010gv,Fitzpatrick:2010br,Banks:2010eh}.
The EDM cross section can be calculated from integrating the following expression 
\beq
\frac{d \sigma_{\rm EDM}}{d\cos \theta} & = & \frac{Z^2 e^2}{4 \pi v^2}
\frac{\mu_{\rm el}^2(q^2 (\cos \theta))}{1-\cos \theta} 
\eeq
which is of course divergent for $d =2$, but finite for any
other legitimate choice of~$d$. 

The expression for the MDM cross section is slightly more complicated,
and includes both a finite dipole-dipole piece and a formally divergent
dipole-charge radius interaction. The cross sections for these
interactions are 
\beq
\frac{d\sigma_{\rm MDM-charge}}{d\cos \theta} & = & \frac{e^2
  \mu_{\rm mag}^2(q^2 (\cos \theta)) Z^2}{4 \pi} \left( \frac{1}{1-\cos
    \theta} - \frac{m_\chi}{(m_\chi+m_A)v^2}\right)\\
\frac{d \sigma_{\rm MDM-MDM}}{d\cos \theta} & = & \frac{\mu_Z^2
  \mu_{\rm mag}^2 (q^2(\cos \theta))}{2 \pi v^2}
\eeq
One can get total cross sections for the DM-nucleon scattering from
integrating these expressions over $\cos \theta$. However, as we will
explain later in Sec~\ref{subsec:YXsec}, the direct detection
experiments are sensitive to the effective cross sections, rather than
``honest'' total cross sections. Therefore, we will return to this
question in Sec.~\ref{subsec:YXsec} and get numerically the effective
cross sections, taking into account the constraints on
operator~\eqref{eq:Ycoupling}. 

\subsection{The BBN constraint}
\label{subsec:hyperBBN}

As we have mentioned in Section \ref{subsec:BBNconcern}, we imagine
that there is a mass gap for particles created by the operator ${\cal
  O}_{\mu \nu}$ and that these particles decay before BBN. First, let
us suppose that the {\em only} particles we need to concern ourselves
with are those created by ${\cal O}_{\mu \nu}$. Consider a vector
particle $V$ of mass $m_V$, which has a matrix element 
\be
\left< 0 \middle| {\cal O}_{\mu \nu}(x) \middle| p, \epsilon\right> = \xi_V m_V^{d - 2} \left(\epsilon_\mu p_\nu - \epsilon_\nu p_\mu\right) e^{i p \cdot x}. \label{eq:matrixelementV}
\ee
Here $\xi_V$ is an order-one constant whose value can only be computed if we understand the detailed dynamics behind the mass gap.

Through the mixing of ${\cal O}_{\mu \nu}$ with hypercharge, the $V$ particle can decay into charged Standard Model particles. In the mass range $m_V \sim 10~{\rm MeV}$ that is primarily of interest to us, the only kinematically accessible particles will be electrons since $m_{e^-} \approx 511~{\rm keV}$. Neutrinos also carry hypercharge, but we are considering a process at energies below the $Z$ boson mass for which a decay to neutrinos will carry extra $m_V^2/m_Z^2$ suppression in the amplitude. As a result, we are interested in computing the partial width for the decay $V \to e^+ e^-$. This is closely analogous to the $\rho^0 \to e^+ e^-$ process in QCD, which proceeds via kinetic mixing of the rho and the photon, or the frequently studied case of dark photons (e.g.~\cite{Pospelov:2008zw}). The kinetic mixing of the $V$ particle is 
\be
\frac{\epsilon_V}{2} V_{\mu \nu} F^{\mu \nu}, \quad \epsilon_V = 2 c_B \xi_V \cos \theta_W \left(\frac{m_V}{\Lambda}\right)^{d-2}.
\ee
We compute
\be
\Gamma(V \to e^+ e^-) \approx \frac{4}{3} \xi_V^2 c_B^2 \alpha \cos^2 \theta_W  m_V \left(\frac{m_V}{\Lambda}\right)^{2d-4}\left(1 + \frac{2m_e^2}{m_V^2}\right) \sqrt{1 - \frac{4 m_e^2}{m_V^2}}. \label{eq:GammaV}
\ee
We require the lifetime $\tau(V \to e^+ e^-) \simlt 1~{\rm s}$ in order to avoid BBN constraints. This translates into a bound (dropping the phase space factors, which are negligible if $m_V \gg 1~{\rm MeV}$):
\be
\xi_V^2 c_B^2 \left(\frac{m_V}{\Lambda}\right)^{2d-4} \left(\frac{m_V}{10~{\rm MeV}}\right) \simgt 9 \times 10^{-21}. \label{eq:BBNvector}
\ee
As expected, then, decay of the vector can easily happen before BBN when $d$ is not too large. Alternatively, we can express the bound in terms of the effective mixing as
\be
\epsilon_V \simgt 1.7 \times 10^{-10} \, \sqrt{\frac{10~{\rm MeV}}{m_V}}. \label{eq:BBNvectormixing}
\ee
We note that this estimate could be overly conservative: more detailed computations of constraints on dark photons from BBN and the CMB have found that smaller values of $\epsilon_V$ may be safe \cite{Fradette:2014sza}. On the other hand, these studies have assumed a single dark photon with abundance determined from thermal equilibrium; in our case, there is a whole dark sector. Precise constraints may depend on the full model of the sector, but we will take our simple estimates to be the best guide available in the absence of a detailed model-dependent study.

The above constraint is the correct one if the only particles in the approximately scale-invariant sector are vectors, in which case they all decay directly through the mixing with hypercharge. On the other hand, familiarity with a range of examples of strongly-interacting theories tells us that often the lightest states are scalars or pseudoscalars. These could decay through other portals, like the Higgs portal---though as we noted above, the small value of the electron Yukawa suppresses decays through that portal. Alternatively, we could assume that the mixing between ${\cal O}_{\mu \nu}$ and hypercharge is the only coupling between this sector and the Standard Model. It does not permit a direct decay of a light scalar into Standard Model states. However, the scale-invariant sector will in general admit couplings of such a light scalar to heavier vectors, which in turn mix with the Standard Model. Suppose, then, that we have in addition to the vector $V$ considered above a scalar $S$ with mass $m_S < m_V$. We expect that the strong sector will contain couplings like
\be
{\cal L}_{\rm eff} = \frac{\xi_{SVV}}{m_V} SV_{\mu \nu}V^{\mu \nu},
\ee
where we assume that $m_S \sim m_V \sim m_{\rm gap}$ are all of the same order and $\xi_{SVV}$ is an order-one number.  This effective coupling allows the decay $S \to \gamma \gamma$ through two insertions of $V-\gamma$ mixing. We compute this decay width to be:
\be
\Gamma(S \to \gamma \gamma) \approx \xi_{SVV}^2 \xi_V^4 \cos^4 \theta_W c_B^4 \left(\frac{m_V}{\Lambda}\right)^{4 d - 8} \frac{m_S^3}{4\pi m_V^2}. \label{eq:GammaS}
\ee
Again, if we impose a $\tau \simlt 1~{\rm s}$ constraint this leads to a bound:
\be
\xi_{SVV}^2 \xi_V^4 c_B^4 \left(\frac{m_V}{\Lambda}\right)^{4 d - 8} \left(\frac{m_S}{10~{\rm MeV}}\right)^3 \left(\frac{15~{\rm MeV}}{m_V}\right)^2 \simgt 3.2 \times 10^{-21}. \label{eq:BBNscalar}
\ee
Again, we can write this as a bound on the mixing
\be
\epsilon_V \simgt 1.3 \times 10^{-5} \, \xi_{SVV}^{-1/2} \left(\frac{10~{\rm MeV}}{m_S}\right)^{3/4} \left(\frac{m_V}{15~{\rm MeV}}\right)^{1/2}. \label{eq:BBNscalarmixing}
\ee
In the alternative case that the lightest state is a pseudoscalar $P$, the logic is essentially identical with the modification that the effective coupling is of the type $P V_{\mu \nu} {\tilde V}^{\mu \nu}$. We take this to be a conservative bound on the constraint imposed by BBN on the hypercharge portal. Of course, if the lightest state really is a vector, we can use the safer bound from $V \to e^+ e^-$. 

In both cases, we see that the most obvious danger arises when $d \gg 2$. In that case the small ratio $m_V/\Lambda$ is raised to a significant power and could potentially become small enough to cause problems. On the other hand, direct constraints tend to be strongest at small values of $d$, so we will carefully check whether the parameter space we consider below conflicts with these constraints. However, as we noted in Section \ref{subsec:BBNconcern}, it is also possible that a nonstandard cosmological history can eliminate the problem even if these bounds are violated.

\subsection{Further constraints from dark photon bounds}
\label{subsec:darkphoton}

Many experimental searches have placed constraints on a massive vector particle $V$ whose field strength mixes with that of hypercharge. These are generally referred to as ``dark photon'' searches and rely on a combination of precision measurements, searches for rare meson decays, and low-energy collider or fixed target experiments \cite{Pospelov:2008zw,Batell:2009yf,Reece:2009un,Bjorken:2009mm}. Further constraints on the small-mixing regime arise from the effect of dark photon production on energy loss in supernovas \cite{Kolb:1996pa}. Recent reviews of the status may be found in \cite{Jaeckel:2013ija,Essig:2013lka}. The continuum that we are interested in is a limiting case of a tower of dark photons with different masses and with a mass-dependent  kinetic mixing parameter. Some constraints on dark photons carry over directly to our scenario. Others involve searches for narrow spectral features and no longer apply to a continuum of new particles. However, in the presence of a mass gap, it is quite plausible that near threshold there is a well-defined, narrow single-particle state. In the case that the spectral function consists of a sum over narrow poles from a warped extra dimension, it has been argued that the lightest state will give rise to the most important constraint \cite{McDonald:2010iq}.

One of the strongest constraints for resonances in the mass range of interest, near 10 MeV, is from the NA48/2 experiment at CERN, which searched for decays $\pi^0 \to \gamma V$ with $V \to e^+ e^-$ \cite{Batley:2015lha}. They have expressed the constraint as an upper bound on $\epsilon_V^2$ which fluctuates, over the mass range $10~{\rm MeV} \simlt m_V \simlt 20~{\rm MeV}$, between about $2 \times 10^{-7}$ and $8 \times 10^{-7}$. Other interesting bounds come from beam dump experiments \cite{Bjorken:2009mm,Andreas:2012mt,Blumlein:2013cua}. For $\epsilon_V \sim 10^{-3}$, the Fermilab E774 experiment \cite{Bross:1989mp} excludes $m_V \simlt 10~{\rm MeV}$. For masses $10~{\rm MeV} \simlt m_V \simlt 20~{\rm MeV}$, a combination of the SLAC E141 \cite{Riordan:1987aw}, SLAC E137 \cite{Bjorken:1988as}, KEK \cite{Konaka:1986cb}, and Orsay \cite{Davier:1989wz} beam dump experiments exclude a wide range of mixings between about $10^{-8}$ and ${\rm few} \times 10^{-4}$. Supernova constraints further exclude the range $10^{-10} \simlt \epsilon_V \simlt 10^{-8}$ for dark photon masses up to around 100 MeV \cite{Kolb:1996pa,Dent:2012mx,Dreiner:2013mua} and the lack of observations related to electromagnetic decays of the dark photon outside the supernova exclude a further range $10^{-12} \simlt \epsilon_V \simlt 10^{-10}$ for $m_V \simlt 20~{\rm MeV}$ \cite{Kazanas:2014mca}.

\begin{figure}[t]
\centering
\includegraphics[width=.5\textwidth]{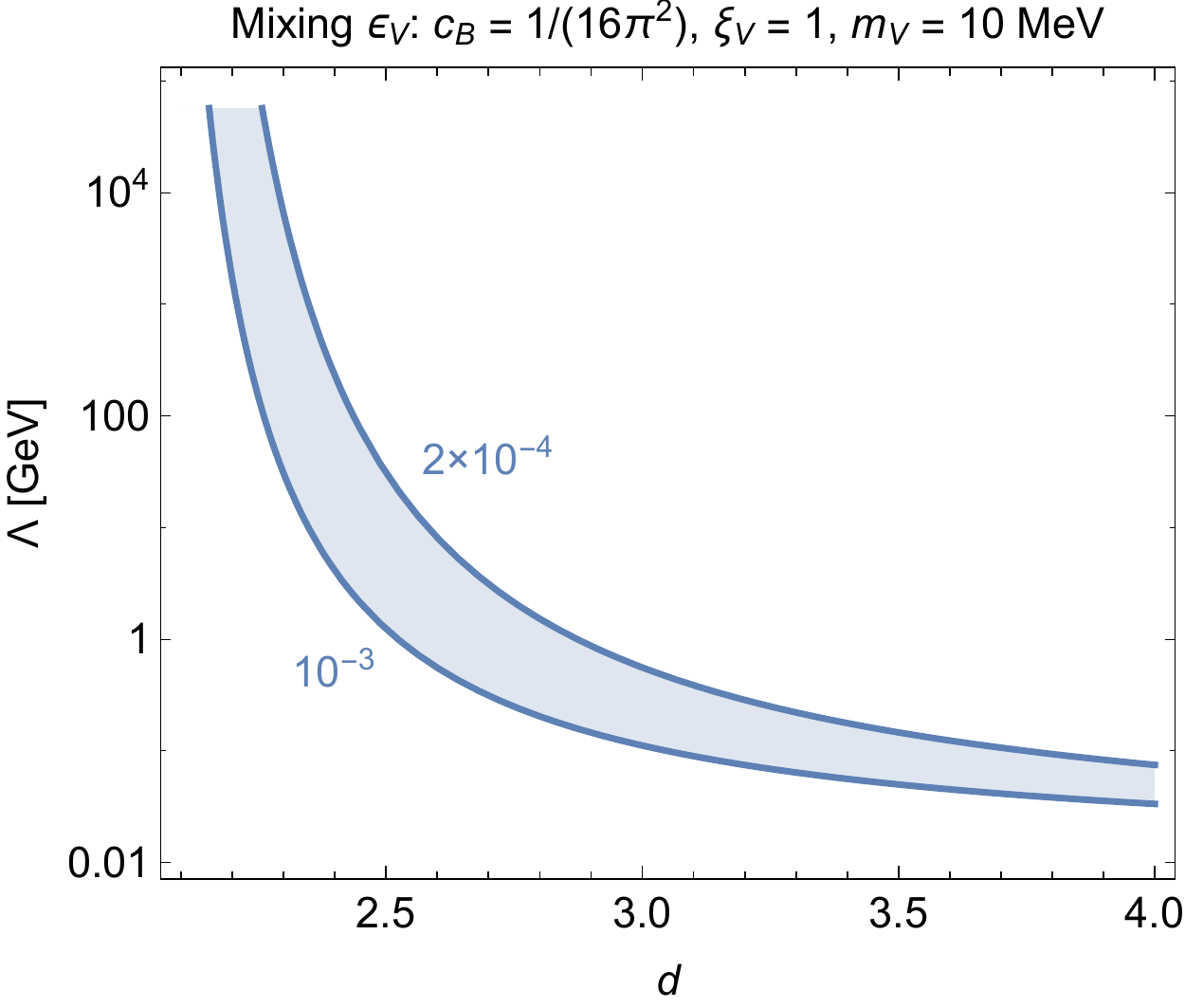}
\caption{Region of parameter space allowed by the inequality (\ref{eq:largewindow}) in blue. We have fixed $c_B = \frac{1}{16\pi^2}$ (representative of a one-loop factor), $\xi_V = 1$, and $m_V = 10$ MeV, and plot the band of allowed mixing $\epsilon_V$ as a function of $d$ and $\Lambda$.}
\label{fig:darkphotonconstraint}
\end{figure}

Reanalyzing all available data for the case of a continuum of states is both beyond the scope of this paper and highly model-dependent. Unlike direct detection phenomenology, searches for dark photons directly probe the timelike region of the two-point function for ${\cal O}_{\mu \nu}$ and the result can be very sensitive to the detailed mechanism for generating a mass gap. Theories with a sequence of many narrow resonances are subject to a wider variety of direct constraints than theories with a broad continuum. We expect that there will likely be a resonance-like state near the mass threshold, based on analogy to QCD-like theories, but even this is not obviously guaranteed by general principles of quantum field theory. Furthermore, the amount of spectral weight in such a state is not guaranteed; that is to say, we might find that the parameter $\xi_V$ in (\ref{eq:matrixelementV}) is significantly smaller than one, relaxing the bounds. Caveats aside, there is essentially only one window in which we might aim for the light state $V$ to lie:
\beq
2 \times 10^{-4} \simlt \epsilon_V \simlt 10^{-3}. \label{eq:largewindow}
\eeq
with $10~{\rm MeV} \simlt m_V \simlt 20~{\rm MeV}$ (to avoid the E774 constraint while still maintaining continuum-like direct detection phenomenology). This window squeezes in between beam dump constraints and the $\pi^0$ decay search. The combination of beam dump and supernova constraints rule out all smaller mixings above the BBN constraint (\ref{eq:BBNvectormixing}) and are fairly robust: for instance, even if further dark sector states exist and a decay like $V \to SS$ is allowed and prompt, the $S$ particle will not promptly decay to Standard Model particles unless couplings through portals other than just hypercharge are added to the theory. In the window (\ref{eq:largewindow}) of allowed $\epsilon_V$, decays of the $V$ particle are prompt on collider length scales and even the BBN constraint (\ref{eq:BBNscalarmixing}) for the case of scalar decays is safe.

This constraints are visualized in Fig.~\ref{fig:darkphotonconstraint}. We see that for reasonable values of the suppression scale $\Lambda$, in the region of allowed $\epsilon$ (in blue) the parameter space $d \to 2$ is compatible with suppression scales a few orders of magnitude above the weak scale. 

\subsection{Constraint from exotic $Z$ decays}

Mixing of the $Z$ boson with the operator ${\cal O}_{\mu \nu}$ allows
the $Z$ to decay into light mediators. These may shower or cascade in various ways, possibly producing a large multiplicity of light particles, like the vector $V$ or scalar $S$ discussed above. Depending on the lifetime of these particles the event could be invisible or there could be distinctive signals involving lepton jets \cite{ArkaniHamed:2008qp}, photon jets \cite{Ellis:2012zp}, displaced vertices, or combinations thereof. Without a detailed model, it is difficult to say precisely what the LEP constraint on such exotic decays might be, or even whether the tightest constraint arises from LEP rather than from a hadron collider. Consequently, we will focus on a very robust limit arising from the total width of the $Z$ boson. The $Z$ boson width is measured to be $\Gamma_Z = 2.4952 \pm 0.0023~{\rm GeV}$ \cite{Agashe:2014kda} whereas the global electroweak fit not including direct measurements of the width gives $\Gamma_Z = 2.4946 \pm 0.0016~{\rm GeV}$~\cite{Baak:2014ora}. From these numbers we estimate the constraint on the {\em total} new physics contribution to the $Z$ boson width,
\be
\Gamma^{\rm new}_Z \simlt 6~{\rm MeV}. \label{eq:Zexoticbound}
\ee
The bound for invisible decays is somewhat stricter, but not dramatically so \cite{Carena:2003aj}.

\begin{figure}[t]
\centering
\includegraphics[width=.9\textwidth]{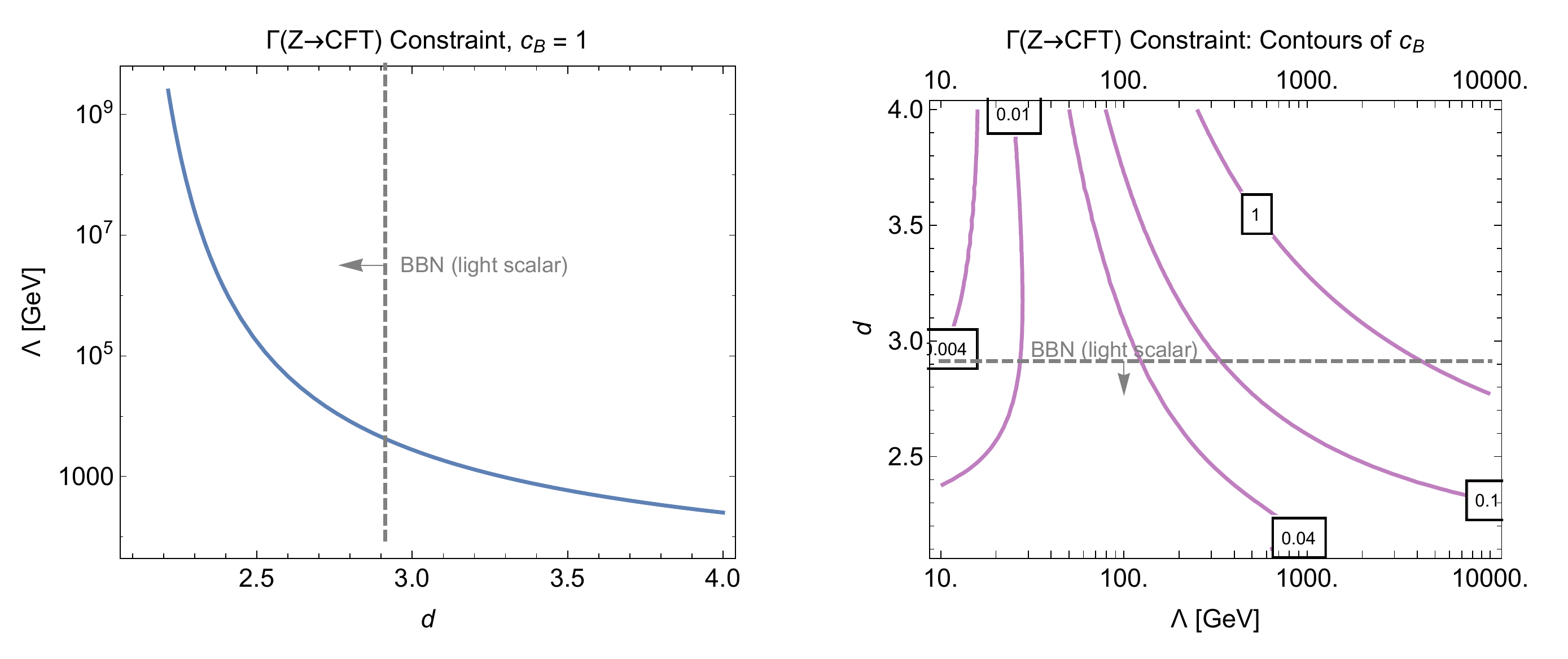}
\caption{Constraints on couplings in the theory from the $Z$ boson
  exotic width. At left, we also fix $c_B = 1$ and plot the largest
  allowed value of $\Lambda$ as a function of the operator dimension
  $d$. In the region to the left of the dashed line, this value of
  $\Lambda$ obeys the constraint (\ref{eq:BBNscalar}) for BBN with a
  10 MeV scalar as the lightest state; the entire plot obeys the
  constraint (\ref{eq:BBNvector}) for BBN with a 10 MeV vector
  lightest particle. At right, we vary $\Lambda$ and $d$ and plot
  contours of the largest allowed value of $c_B$. Again, the dashed
  line demarcates the part of the plot that is safe from BBN even with
  a light scalar.} 
\label{fig:Zwidthconstraint}
\end{figure}

We will denote the decay width induced by the mixing of the $Z$ with ${\cal O}_{\mu \nu}$ by $\Gamma(Z \to {\rm CFT})$, even though a small mass gap may mean that strictly speaking we are not decaying to a conformal sector. To compute this width we exploit the optical theorem, which tells us that $\Gamma(Z \to {\rm CFT}) = \frac{1}{m_Z} {\rm Im}{\cal M}(Z \to Z)$, where ${\cal M}(Z \to Z)$ is the amplitude for the $Z$ boson to mix with the mediator and then mix back. The imaginary part of the amplitude comes from the discontinuity across the branch cut in the factor $(-k^2)^{d-2}$ in the two-point function of ${\cal O}_{\mu \nu}$: ${\rm Disc}\left[(-k^2)^{d-2}\right] = 2 i \sin(\pi d)(-k^2)^{d-2}$. Using the identity $\sin(\pi d) = \frac{\pi}{\Gamma(d)\Gamma(1-d)}$, we find
\be
\Gamma(Z \to {\rm CFT}) = \frac{2 \pi  c_B^2 \sin^2 \theta_W
  (d-1)(d-2)}{\Gamma(d) \Gamma(d+1)}
\left(\frac{m_Z}{2\Lambda}\right)^{2d-4} m_Z. \label{eq:Zexoticwidth} 
\ee 
Notice that this decay width goes to zero when $d \to 2$; in this
limit the mixing is with a free particle, and the spectral function
has no overlap with the $Z$ mass. Comparing equations
(\ref{eq:Zexoticbound}) and (\ref{eq:Zexoticwidth}) gives us a bound
on the largest allowed mixing for any model. We plot this bound in
Fig.~\ref{fig:Zwidthconstraint}. The figure shows that at small values
of $d$, approaching the limit of simple kinetic mixing, the bounds
become strong and force us to consider large values of $\Lambda$. At
larger operator dimensions, the bound is weaker and even relatively
low values of $\Lambda$ are allowed by the data. These large operator
dimensions ($d \simgt 2.9$) can be in conflict with the BBN constraint
(\ref{eq:BBNscalar}) in the case that the lightest particle in the
mediator sector is a scalar. However, as noted earlier this bound can
be avoided by adding more direct couplings of the scalar operator to
the SM (e.g. to the Higgs portal) or by unconventional thermal
histories of the early universe. 

\subsection{Maximum total cross sections}
\label{subsec:YXsec}

The constraints from dark photon searches and new contributions to the
$Z$ boson width lead to a maximum possible direct detection cross
section for a given operator dimension and 
value of $c_2$. We would like to assess how large a total cross
section it is reasonable to obtain. However, the total cross section
itself is not an ideal quantity to plot. The reason is that,
particularly for models that produce distributions $dR/dE_R$ that are
highly peaked at low energy transfers, the total cross section may not
be reflective of the event rate in a real experiment with a finite
energy threshold. (Indeed, the total cross section may not even be
defined, for sufficiently singular behavior at low $E_R$.) As a
result, we will define a notion of ``effective total cross section.'' 

The typical exclusion plot of an experiment like LUX gives a
constraint on the dark matter--proton scattering cross section. Of
course, this involves an assumption, and makes the most sense for the
case in which the operator involved in the non-relativistic theory is
simply $1$, e.g. when the underlying relativistic interaction is
${\bar \chi}\chi {\bar N} N$. Thus, to get a rough notion of the total
event rate allowed, we take the following definition: 
For a given model of interest, the {\bf effective dark matter--proton
  cross section} $\sigma_p^{\rm eff}$ is defined as the value of
$\sigma_p$ obtained in a model with a 100 GeV Dirac dark matter
particle $\chi$ scattering through the interaction ${\bar \chi}\chi
{\bar p}p$ that gives the {\em same} integrated rate for scattering on
$^{131}$Xe over the recoil energy range 3 keV to 25 keV as the model
of interest. 
This definition has the disadvantage of making explicit reference to a particular experiment and its range of accessible recoil energies. However, {\em any} definition of cross section that we compare across different experiments and different models must build in some assumptions. Rigorously speaking, we should fit each individual model across all experiments. For the simple purpose of getting an order-of-magnitude sense of how the scattering rate allowed by our model compares to the scattering rate of a more standard model, on the other hand, $\sigma_p^{\rm eff}$ is fairly useful.

\begin{figure}[t]
\centering
\includegraphics[width=1.0\textwidth]{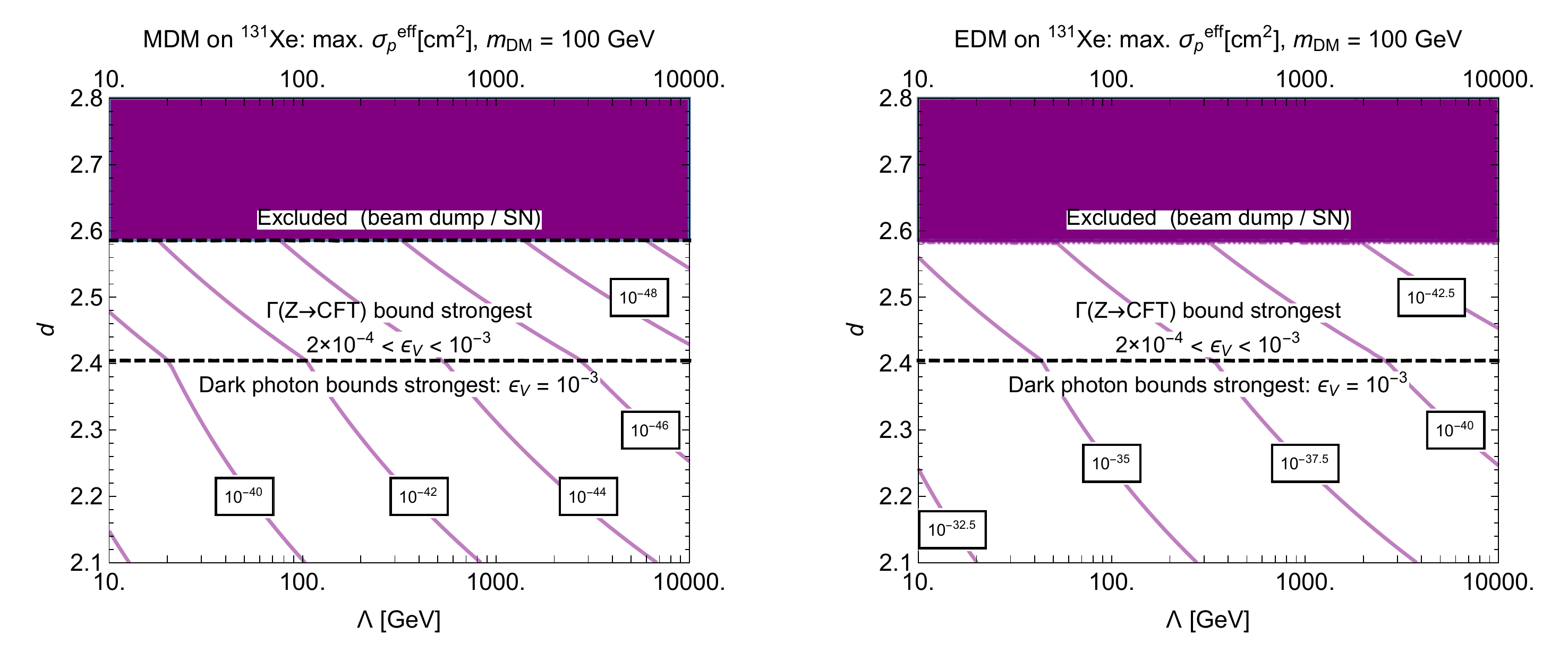}
\caption{The largest effective cross section $\sigma_p^{\rm eff}$ for
  dark matter scattering on $^{131}$Xe allowed by the constraints from
  dark photon searches and the $Z$ boson width. The left-hand plot is
  for magnetic dipole-type couplings of DM to ${\cal O}_{\mu \nu}$ and
  the right-hand plot for electric dipole-type coupling. We fix $c_2 =
  1$ and a dark matter mass of 100 GeV. For a given $\Lambda$ and $d$,
  the largest allowed value of $c_B$ is chosen. At small $d$ this is
  determined by the dark photon constraint $\epsilon_V \simlt
  10^{-3}$. At somewhat larger $d \simgt 2.4$ the $Z$ width becomes
  the dominant constraint. At the point that the $Z$ width no longer
  allows $\epsilon_V \simgt 2 \times 10^{-4}$, the beam dump and supernova
  constraints on dark photons force us all the way down to $\epsilon_V$ so small that the dark photons would decay after BBN, so no cross section is allowed above $d \approx 2.6$. What is plotted is the cross section $\sigma_p$ in square centimeters associated to
  a model with scalar contact interaction achieving the same
  integrated rate.} 
\label{fig:Xeeffectivecrosssection}
\end{figure} 

With these interpretational caveats out of the way, we present the
result for the largest $\sigma_p^{\rm eff}$ allowed by the dark photon
and $Z$ width constraints. We compute $dR/dE_R$ for the comparison
model, ${\cal L} = \frac{f_p}{\Lambda^2} {\bar \chi}\chi {\bar p}p$,
using the code of Ref.~\cite{Anand:2013yka}, then numerically
integrate over $E_R$ from 3 to 25 keV. For this model, $\sigma_p =
\frac{\mu_p^2}{\pi} \left(\frac{f_p}{\Lambda^2}\right)^2$. We also
compute $dR/dE_R$ for our model of interest with the same code,
putting in a contact interaction and then weighting the answer by
appropriate powers of $-q^2$ or of $(-q^2 + m_{\rm gap}^2)$ to model
the continuum exchange. Finally, we scale $f_p$ so that the integrated
rates match and read off the corresponding value of $\sigma_p$. 

The result of this exercise, for dark matter scattering on $^{131}$Xe,
is presented in Fig.~\ref{fig:Xeeffectivecrosssection}. The plot
illustrates that there is no allowed cross
section when $d \simgt 2.6$, because the $Z$ width, beam dump, and supernova constraints forbid all possible $\epsilon_V \simgt 10^{-10}$ and smaller $\epsilon_V$ are excluded by BBN. For smaller values of $d$
compatible with the range $2 \times 10^{-4} \simlt \epsilon_V \simlt
10^{-3}$, the largest allowed cross section is determined mostly by
the dark photon constraint when $d \simlt 2.4$ and mostly by the $Z$
width constraint when $d \simgt 2.4$. Notice that the cross sections
allowed for EDMs are several orders of magnitude larger than for
MDMs. This is because, at fixed coefficient $c_B$, the
non-relativistic scattering through the MDM operator is suppressed by
$v^2 \sim 10^{-6}$ relative to that through the EDM operator. 

To summarize our results, despite the existence of a variety of
stringent constraints on the operator ${\cal O}_{\mu \nu}$ coupling to
hypercharge, arising from both low-energy probes like beam-dump
experiments and pion decay and high-energy probes like $Z$ boson
decays, there is a range of dimensions---roughly $2 \simlt d \simlt
2.6$---for which sizable direct detection cross sections could
occur. Of course, direct detection experiments themselves can
constrain this region. However, due to the unusual spectral shapes
that appear in continuum-mediated scattering, a new analysis of the
direct detection data will be necessary to derive precise bounds. We
leave such analyses for future work. 

%% file: conclusions.tex
As the simplest models of WIMPs come under strain from a variety of experiments (direct detection, indirect detection, and colliders), it   becomes increasingly important to broaden our theoretical vision of what dark matter might be. In recent years, a large number of directions have been explored. One common theme is the possibility of a dark sector, with additional new particles beyond the dark matter alone. In this paper we have explored a novel type of dark sector in which the scattering of dark matter with ordinary matter proceeds not through a contact interaction or the exchange of a single particle but through a continuum of mediators.

The basic formalism for direct detection is simple: the continuum mediator multiplies the amplitude by a function of $q^2$, which in the simplest scale-invariant scenario is just a power law. However, as we have seen, various complications arise. If we wish to have a mass gap to avoid BBN constraints, the direct detection phenomenology can be surprisingly robust, but the constraints imposed on the scenario from other experiments may be very sensitive to the nature of the mass gap. We have explored this in some detail for the case of an antisymmetric tensor mediator coupling to the field strength of hypercharge. A combination of low-energy and high-energy accelerator experiments puts significant restrictions on the allowed parameter space. Nonetheless, we have argued that these constraints still allow room for quite large and detectable signals at direct detection experiments like LUX.

Many aspects of our analysis could be refined in the future, as the interplay between direct detection experiments and other experiments is complex. We have not discussed indirect detection, where the Sommerfeld effect may be relevant (see \cite{Chen:2009ch} for related work). The annihilations are expected to proceed dominantly into the states of the approximately conformal sector, 
undergo complicated cascades and then finally decays into the SM particles. Eventually the annihilation mode closely resembles 
the hidden valley scenario ($2\to {\rm many}$). This distinctive annihilation pattern together with potential signal in low-energy 
searches for the light dark photon can potentially help to differintiate this scenario from an ordinary form-factor DM, which otherwise looks identical to the continuum mediated DM scenario in the direct detection experiments.    
These directions would be interesting to explore further. 

Despite these complexities, the picture for direct detection is extremely simple, and searches for unusual dependence of signals on recoil energy are well worth pursuing. It is an exciting possibility that the discovery of dark matter could also be the discovery of a rich, interacting sector that could exhibit novel quantum field theoretic phenomena like scale-invariance in ways that we have not previously seen in particle physics.